# Universal Metallic Surface States in Electride

Dan Wang,[1,2] Hongxing Song,[2] Leilei Zhang,[3] Hao Wang,[2] Yi Sun,[2] Fengchao Wu,[2] Ying Chen,[4] Xiangrong Chen,[1*] Hua Y. Geng[2,5*]

[1] *Institute of Atomic and Molecular Physics, College of Physics, Sichuan University, Chengdu 610065, P. R. China;*

[2] *National Key Laboratory of Shock Wave and Detonation Physics, Institute of Fluid Physics, China Academy of Engineering Physics, Mianyang, Sichuan 621900, P. R. China;*

[3] *Institute of Nano-Structured Functional Materials, Huanghe Science and Technology College, Zhengzhou 450063, P. R. China;*

[4] *Fracture and Reliability Research Institute, School of Engineering, Tohoku University, Sendai 980-8579, Japan;*

[5] *HEDPS, Center for Applied Physics and Technology, and College of Engineering, Peking University, Beijing 100871, P. R. China.*

**SIGNIFICANCE STATEMENT:**

Metallic surface states (MSS) are important in broad applications such as chemical catalysis and quantum computing. Besides topological insulators, trivial material also could host MSS occasionally. They all are, however, very fragile. Searching for robust MSS is thus a holy grail. Here we demonstrate that electrides, even might be classified as topologically trivial by standard theory, always possess robust and universal MSS against any disturbances. This fact not only uncovers the limitation of the standard topological theory, but also recognizes insulating electrides as a conceptually new type of topology beyond the standard theory, and opens up new opportunity to MSS with large bulk gap.

* To whom correspondence should be addressed. E-mail: s102genghy@caep.cn; xrchen@scu.edu.cn.

## ABSTRACT

Robust metallic surface states (MSS) of topological insulator (TI) against imperfections and perturbations are important in broad applications such as chemical catalysis and quantum computing. Unfortunately, they are suffered from the narrow band gap that can be accessed. Searching for MSS with large bulk band gap beyond conventional TIs becomes a quest. In this work, inspired by the adiabatic connection principle in real space, we identify that all electrides, a new class of emerging materials, must host robust and universal MSS that resists any disturbances, in spite of the fact that some of them could be classified as trivial in standard topology theory. This counterintuitive property is traced to the specific charge localization-delocalization change intrinsic to electride when approaching the crystalline surface or interface, which is a kind of interstice-centered to atom-centered transition in the real-space topology of the charge density distribution, and is sharply different from the band inversion in the standard topology theory. The new mechanism circumvents the obstacle that limits the band gap of TI. Robust and universal MSS in an electride that conventionally-determined as trivial but with a colossal band gap beyond 6.13 eV are demonstrated. This gap size is about 6-fold larger than the highest record of known "wide-gap" TIs, thus opens up new avenues to universal MSS with gigantic bulk gap.

# 1 Introduction

Topological insulators (TI) and topological semimetals with metallic surface states (MSS) or edge states have attracted great interest recently[1, 2]. A variety of topologically nontrivial materials have been predicted by theory and confirmed by experiments, such as Bi-based materials[3-6]. They are, however, suffered from narrow bulk band gap, with a typical size less than 0.2~0.7 eV[7-9], even for the touted "wide-gap" as 1.0 eV in some Bi/Sb halides and hydrides[10, 11]. This greatly hinders their practical applications. Searching for robust MSS with large band gap therefore becomes a holy quest in this field[8, 12]. Conceptually novel state beyond the standard topology definition, such as obstructed atomic insulator (OAI)[9], has been proposed. Unfortunately, the MSS of OAI can occur only on certain specific cleavage plane, thus is unstable against stacking faults or poly-crystallization.

On the other hand, recent high-pressure studies have indicated that some electrides might exhibit MSS[13]. This was thought as originated from the band inversion like in conventional TIs. Nonetheless, it also could arise from a generic property of all electrides, which are characterized by highly-localized electrons in the interstices of the lattice that serve as anions and known as interstitial quasi-atom (ISQ)[14]. Based on the dimensionality of the anionic electrons and interstitial voids, electrides can be classified into zero-dimensional (0D), one-dimensional (1D), two-dimensional (2D), and three-dimensional (3D)[15, 16]. They all have an extraordinary topology of interstice-centered charge density distribution, which is quite different from the normal atom-centered distribution. Despite this distinct topology difference in real space, it was believed that though some electrides could be topologically non-trivial[17-20], many of them should be topologically trivial[21] in terms of standard topological band theory (TBT).

Furthermore, it was also known that the TBT formulated in reciprocal space is not effective to exhaust all possible topological phases. For example, only fewer than 400 topological materials were identified out of 200000 existent in crystal structure database by the reciprocal space band theory[22]. However, with the same database, many more topologically non-trivial materials were screened out by the real-space theory of topological quantum chemistry (TQC)[6, 22, 23]. In addition, according to the TQC[9], topologically trivial insulators can be further classified into atomic insulators and OAI. The former has the Wannier orbital centers at the atomic sites, whereas for OAI at least one of the Wannier centers must locate at empty sites, leading to MSS on special cleavage planes on which no atoms co-planar with the given Wannier centers. It is necessary to note that the MSS due to topological requirement do not necessarily connect the valence bands and the conduction bands[23, 24]. Conventional TBT and TQC all are built upon lattice symmetry-group argument and band representation[22]. They cannot capture other topological changes outside this realm. For example, there might have other topological ordering in real space that cannot be well represented by quasiparticles or their eigenvalues.

On the other hand, it is surprise that alkali metals under high pressure might exhibit a wide diversity of MSS which closely relate to localized charge transfer[25]. These localized orbitals usually behave like the spatially extended $s$-electrons with a spherical symmetry and off the atomic centers. Delocalization and overlapping of these localized orbitals can lead to gapless state[25]. In this sense, insulating electrides might host MSS if the ISQs near the crystalline surface are partly damaged and delocalized. Like OAI, this mechanism circumvents the direct requirement of band inversion as in conventional TIs[7], thus is promising to open a new playground for MSS with large bulk band gap. It also should note that electrides with MSS could promote wide diverse applications, such as catalysts[26-29], novel battery electrode materials[30-32], electron emitters[33-36], superconductors[37-40], magnetoelectric materials[41], quantum dots[42], quantum computing and spintronic devices[43-45]. In particular, spin-polarized catalysts and electron emitters could be envisioned to emerge in a topological electride, where strong charge polarization could provide versatile capability to tune the property of materials[46, 47].

To this end, we first need to answer following questions: do all insulating electrides have MSS? If yes, why? What symmetries and surface terminations are required to preserve the MSS? The answer to these questions could inspire the design of novel MSS and promote practical application of electrides.

So far, the analysis of MSS and topology is usually based on the electronic band structure in reciprocal space[9, 22], which though is elegant and powerful, sometimes might be less intuitive. In this work, we attempt to address these fundamental questions in real space by tracing the evolution of the localized electrons (or ISQ) from the bulk interior of an electride to its surface. The adiabatic connection approach[48] is employed to illustrate the emerging of MSS. After demonstrating the ubiquitous existence of MSS in typical electrides, we establish a general conclusion on the evolution of ISQ in the vicinity of surface or interface. That is, due to the fact that electrides have a topologically different distribution of localized charge from a normal material, the resultant surface or interface states must be gapless because of the unavoidable overlapping and delocalization of the interstitial charge by this topology transition. With this knowledge, we further predict that all insulating or semimetallic electrides must host a MSS.

## 2 Methodology and computational details

### 2.1 Adiabatic connection method

Adiabatic connection is an approach that is widely used in theoretical physics to establish physical relation between systems, to decouple interactions, to make approximations, or to construct a theoretical formalism. For example, the well-known Kohn-Sham formalism of DFT is constructed with the adiabatic connection method by mapping an interacting system onto a non-interacting system[49]. In electrides, electrons are localized to the interstitial sites, whereas at the outside of the surface, all of them should distribute around the atoms (*i.e.*, atom-centered)[42]. It is rational to assume an adiabatic variation of the charge density distribution evolving from the interior of the bulk to the outside of the surface (or beyond the interface). Considering that all physical processes actually take place in real space, the real space variation with adiabatic connection becomes a natural choice to demonstrate the emergence of novel surface states in electrides that arising from the heterogeneity induced by the termination of lattice symmetry and periodicity, as well as the change in charge density distribution topology.

We formulate the real space adiabatic connection in electrides as follows. Assume the Hamiltonian of an electride can be fully described by the electronic and atomic coordinates in real space, namely $H \equiv H(\boldsymbol{r})$, then we can construct a fully localized version of this Hamiltonian with, say, a tight-binding representation. In order to establish an adiabatic connection, we need to separate and divide this localized Hamiltonian according to the index of each unit cell along the path that links the bulk interior to the outside of the surface. For that purpose, we take out a unit cell that labelled as $j$ along the path, and impose periodic boundary conditions on it. The resultant Hamiltonian is written as $H_{lj} = H_l(\boldsymbol{r}_j) + \Delta_j$, in which the subscript $l$ indicates local tight-binding Hamiltonian, and $\Delta_j$ represents an auxiliary field to ensure that $H_{lj}$ equivalent to the original localized Hamiltonian at point $j$ along the path. In this representation, the local Hamiltonians $\{H_{lj}\}$ give the exact variation of local eigenvalues and eigenstates along the connection path as the original Hamiltonian $H$.

Let us define the adiabatic connection strength of coupling $\lambda$, which continuously varying from $\lambda = 0$ at the interior of a bulk to $\lambda = 1$ at the limit with atom-centered charge distribution. This allows us to relabel $H_{lj}$ as $H_{l\lambda}$ and establishes a generic adiabatic connection path (ACP) from the bulk of an electride to its surface. Within the bulk, $\lambda = 0$, $H_{l0} = H_l(\boldsymbol{r}_0) + \Delta_0$; at the surface or interface of the material, $\lambda = \lambda^*$, $H_{l\lambda^*} = H_l(\boldsymbol{r}_{\lambda^*}) + \Delta_{\lambda^*}$; and at the atom-centered limit outside the surface or interface, $\lambda = 1$, $H_{l1} = H_l(\boldsymbol{r}_1) + \Delta_1$. The local systems along the ACP are thus represented by a series of local Hamiltonian of $H_{l\lambda} = H_l(\boldsymbol{r}_\lambda) + \Delta_\lambda$, for $\lambda = 0 \rightarrow 1$. Their eigenvalues and eigenstates describe how the local electronic state evolves from the bulk state to the atom-centered limit, via the intermediate surface or interface state.

With the eigenstates $\left|\psi_{l\lambda}^{i}\right\rangle$ that associated with an eigenvalue $E_{l\lambda}^{i}$ that satisfy $H_{l\lambda}\left|\psi_{l\lambda}^{i}\right\rangle = E_{l\lambda}^{i}\left|\psi_{l\lambda}^{i}\right\rangle$, we have the local charge density as $n_{l\lambda}(\boldsymbol{r}_\lambda) = \sum_i \psi_{l\lambda}^{i}{}^{*}(\boldsymbol{r}_\lambda) \cdot \psi_{l\lambda}^{i}(\boldsymbol{r}_\lambda)$. The allowed variation of $n_{l\lambda}$ along the ACP for different types of materials is summarized as below:

(*i*) Ionic compounds: all electrons are always localized to the nuclei and being atom-centered, so that $n_{l\lambda}\,(\lambda\colon 0 \to 1)$ changes little, and the charge density always corresponds to a superposition of cations and anions.

(*ii*) Metals: part of electrons are localized to nuclei (*i.e.*, the cores), and others are fully delocalized and shared by all atoms; Therefore $n_{l\lambda} = n_{l\lambda}^{core} + \Delta_{l\lambda}^{share}$. In this case, the charge distribution maintains its main topology when $\lambda$ varying from 0 to 1.

(*iii*) Covalent compounds: electron clouds overlap at the bond center, but the total charge density is still localized to atomic centers[42], so that $n_{l\lambda}$ has little change when $\lambda = 0 \to 1$.

(*iv*) Electrides: A part of electrons are localized to nuclei, but other part is fully localized to the interstitial sites (i.e., off-atomic centers), so that $n_{l\lambda} = n_{l\lambda}^{core} + \Delta_{l\lambda}^{off}$ when $\lambda = 0$. On the outside of the surface, there must be a significant change in charge density distribution to reach the atom-centered limit for all electrons at $\lambda = 1$.

Generally, metallic states must be developed from global charge delocalization[50]. Similarly, MSS could occur if there is heterogenous but global electron delocalization along the ACP. Since only electrides could have prominent change in the charge density (in its real space topology), we consider only the case (*iv*) hereinafter. Namely, we will trace the variation of local charge density distribution $n_{l\lambda}$ from $\lambda = 0$ to $\lambda = 1$ to explore the possible MSS in electrides. In practice, we do not need to know the explicit form of $H_{l\lambda}$. Instead, we could map $H_{l\lambda}$ onto a slab model to bypass the requirement to calculate the local auxiliary fields. Within this approximation, $n_{l\lambda}$ can be calculated with first-principles method directly.

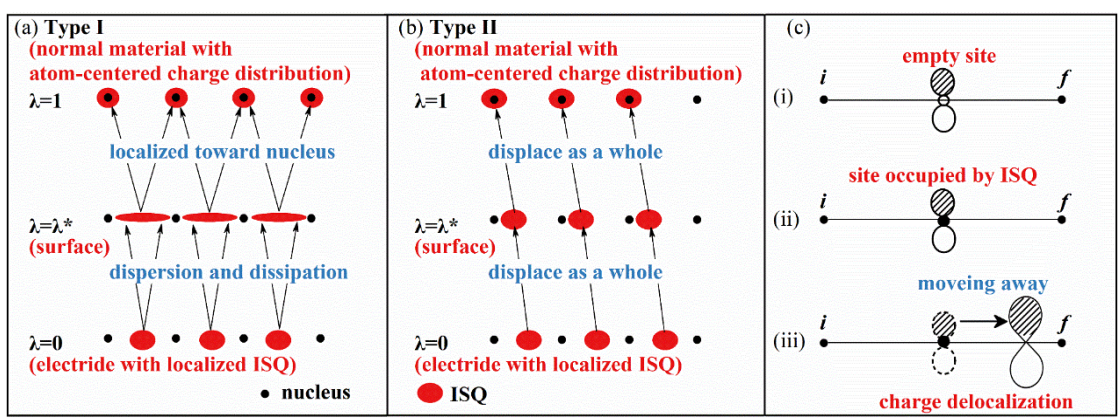

**Figure 1.** (Color online) (a-b) The schematic model that illustrates two possible routes for transferring of localized ISQ to atom-centered limit along the adiabatic connection path (ACP): Type I involves global charge delocalization, whereas in Type II the ISQ moves as a whole and no delocalization involved; (c) Evolution of a set of interstice-centered orbitals in an adiabatic process along a high-symmetry line: in (i) the orbitals at the empty site are gotten stuck by symmetry, this case corresponds to OAI; in (ii) the orbitals and ISQ at the empty site are gotten stuck by symmetry, and corresponds to OAI of electride; in (iii) the orbitals and ISQ at the empty site are not fixed by the symmetry, however, when they move away from the preferred site, the charge of ISQ delocalizes due to the intrinsic nature of electride, and corresponds to the mechanism depicted in (a). In (c), the point *i* and *f* denote high-symmetry sites, which usually are occupied by atoms.

Logically, there are only two types of possible charge transfer route from the localized interstitial sites to the atom-centered limit outside the surface of an electride, as illustrated in Fig. 1(a-b). Type I involves a global delocalization-relocalization process, and thus must lead to a gapless state at or near the surface. By contrast, the ISQ moves as a whole in the Type II, and no delocalization or gapless state involved. The key point here is that in electrides when along the ACP from $\lambda = 0 \to 1$ whether $\Delta_{l\lambda}^{off}$ (*i.e.*, the ISQ) can move as a whole from the interstitial sites to the atomic nuclei or not? Preliminary assessment indicates that this route (Type II) is not favored in energetics, since ISQ in electride is closely related to multi-center bonding[42, 51, 52], the energy cost is high if it moves as a whole[21]. On the other hand, charge delocalization (*i.e.*, via dispersion and dissipation of ISQ) is energetically much easier to achieve [see (iii) in Fig.1(c)].

In next sections, we will examine this argument in typical electrides by tracing the variation of localized charge distribution (as well as the valence state of atoms and ISQs) from the bulk to the surface using a slab model to approximate the ACP.

## 2.2 Computational details

First-principles calculations are performed by using the density functional theory (DFT)[53, 54] with the projector augmented wave (PAW)[55] method and the Perdew-Burke-Ernzerhof (PBE) parameterization of the generalized gradient approximation (GGA)[56] functional as implemented in the Vienna Ab-initio Simulation Package (VASP)[57, 58]. The energy cutoff for the plane wave basis set is chosen as 700 eV in all calculations. The largest spacing between k-points is set as 0.1 Å$^{-1}$ to generate the automatic k-point meshes for the Brillouin zone sampling. All calculations are guaranteed to converge to $10^{-5}$ eV/atom for energy and 0.01 eV/Å for the Hellmann-Feynman forces acting on each atom, respectively. The $1s^22s^1$, $2s^22p^63s^1$, $3p^64s^2$, $6s^26p^2$, $5s^25p^66s^1$, $2s^22p^4$, $3s^23p^63d^14s^2$ and $2s^22p^2$ electrons are treated as valence electrons for Li, Na, Ca, Pb, Cs, O, Sc and C atoms, respectively. The Bader charge analysis[59] is employed to characterize the variation of the valence state of atoms and ISQ. A slab model with a thickness of at least 8-unit cell film is built to simulate the ACP, in which a vacuum gap of 20 Å is used.

## 3 Results and discussion

### 3.1 Robust MSS in insulating dense sodium

Previous reports have found that sodium becomes an insulating electride under high-pressure[60], in which a pressure-induced transformation into an optically transparent phase (Na-hP4) was experimentally observed. Theoretical calculation identified Na-hP4 as an insulting phase with a band gap of 3.5 eV at 320 GPa[60]. The insulating bulk state is formed because compression causes the *3d* bands to rapidly drop in energy relative to the *3p* bands, and increase the hybridization between them[61]. The localized anionic electrons (ISQ) play a key role in this metal-insulator transition. As shown in Fig. 2(a), the main peaks of the density of states (DOS) in the vicinity of the Fermi level primarily consist of ISQs, corroborating the fact that the insulating state in the bulk arises from the strong localization of the interstitial electrons.

In previous investigations, Na-hP4 was thought being topological[25]. However, with the standard TBT or TQC, this phase is determined to be topological trivial; it is even not an OAI if without spin-orbital coupling (SOC), according to the Materials Flatband Database[9]. Thus, it should not have MSS, or the MSS should not be universal and must be fragile if it presents by accident. Unexpectedly, we find MSS on all possible lattice terminations and cleavage planes in this phase for with and without SOC, not just the specified $(10\bar{1}1)$ plane of OAI when SOC is turned on. For clarity, we will focus the discussion on (0001) plane without SOC thereinafter. Other lattice orientations give similar results.

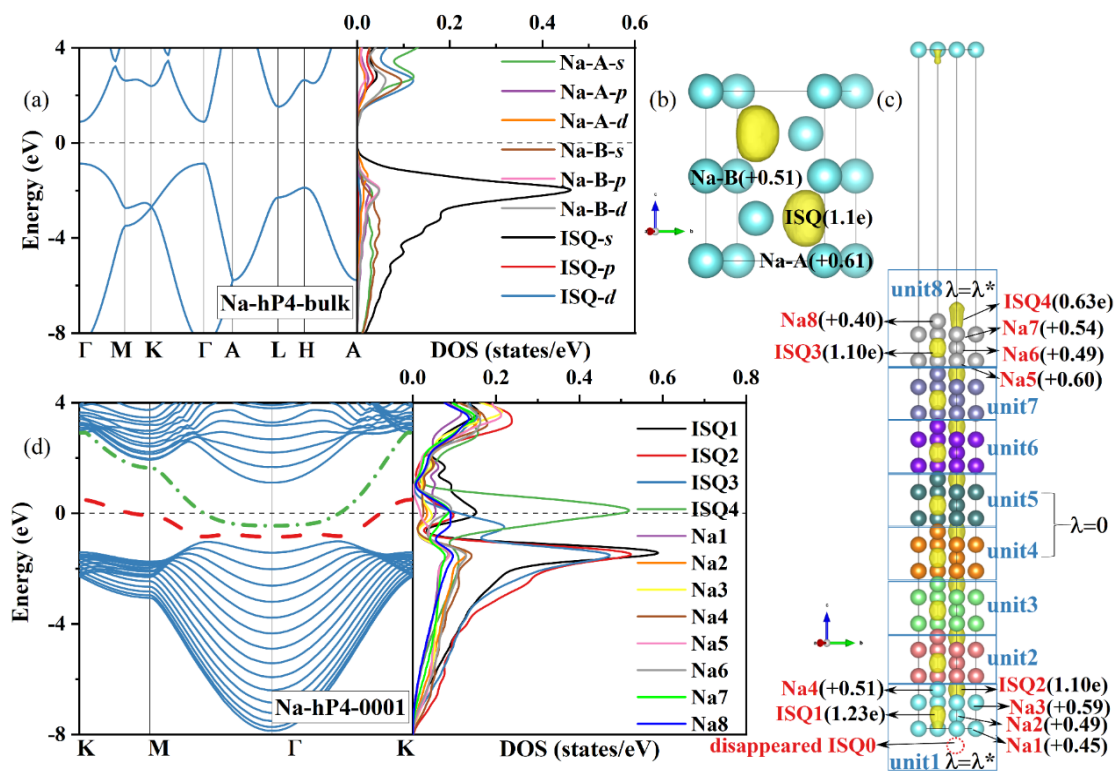

**Figure 2.** (Color online) (a) Band structure (left panel), projected density of states (PDOS; right panel) of Na-hP4 at 320 GPa; (b) Electron localization function (ELF) of Na-hP4, (isosurface=0.9); (c) ELF of the (0001) slab model of Na-hP4, (isosurface=0.75); (d) Band structure (left panel) and PDOS (right panel) of Na-hP4 in (0001) slab model, in which the red (dashed line) and the green (dash-dotted line) surface bands across the Fermi level are highlighted.

For the slab model stacking along the [0001] direction of the Na-hP4 phase (denoted as Na-hP4-0001), there are Na atoms co-planar with the Wyckoff positions 2*d* which are occupied by ISQ [Fig.2(c)]. This cleavage plane does not meet the condition of OAI. Despite that, the surface state is metallic as shown in Fig. 2(d): two new bands are formed within the gap of the bulk and cross the Fermi level. We find that they are mainly induced by the ISQs near the surface [see Fig.2(c)]. The red band [dashed line in Fig. 2(d)] is mainly contributed from the interstitial electrons of the bottom terminal labeled as ISQ1 and the completely dissipated ISQ0, whereas the partially dissipated ISQ4 located on the top terminal contributes to the most of the green band (dash-dotted line). Since these two lattice terminations are different, the result indicates that the occurrence of MSS is independent of the lattice termination type, but rather on the dispersion of ISQs.

In terms of ACP, as shown in Fig. 2(b), within the bulk of Na-hP4, the interstitial 2*d* sites are occupied by localized electrons, which form anionic ISQs with a Bader charge of 1.11e (*i.e.*, a valence state of -1.11). By contrast, the Na-A and Na-B atoms have a positive valence state of +0.61 and +0.51, respectively. Figure 2(c) shows the evolution of electron localization function (ELF) for a slab model with 32 atomic layers stacking along the [0001] direction. The charge density distribution and electronic properties in the interior of the slab are almost the same as the bulk material. However, there is a significant change in the distribution of local charge density when approaching the surface. The results for a thicker model do not lead to a different conclusion, indicating the convergence of our slab size. Following the adiabatic connection theory, here we mainly focus on the variation of the local charge density distribution, especially those near the top and bottom surface.

A striking phenomenon unveiled in Fig. 2(c) is that there is no signature of a uniform displacement of ISQ as a whole when approaching the surface, as required by the Type II mechanism in Fig. 1. This observation supports the aforementioned assessment, and refutes the possibility of Type II route for electrides. In contrast, due to the influence of the heterogeneity introduced by surfaces, localized ISQs become dispersive and dissipated. At the bottom surface where the corresponding unit cell contains four Na atoms and two ISQs [unit-1 in Fig. 2(c)], the ISQ at the outside of the surface (denoted as ISQ0) becomes completely dissipated, and thus cannot be seen in the ELF of Fig. 2(c). These delocalized charges are then pulled into the inside of the surface, leaving conducting holes behind, and giving rise to new surface band across the Fermi level. On the other hand, in the unit-8 that locates on the top surface of the slab, part of the interstitial electrons (0.30e) are also delocalized. The charge state of ISQ4 decreases from 1.11e in the bulk to 0.63e. This change is partially driven by the strong repulsion between Na atoms and the localized anionic electrons. The observed difference in the charge delocalization between the top and bottom surfaces is originated from the different type of the lattice terminations.

Figure 2 demonstrates that the localized ISQs evolve via delocalization rather than by displacement when approach the surface along the ACP. The dispersion and dissipation are induced by lattice termination that leads to a charge redistribution and delocalization at or near the surface. This is exactly the Type I mechanism as shown in Fig. 1, in which a global delocalization process of $n_{l\lambda}$ is involved, thus the unavoidable metallization. As shown in Fig. 2(d), new bands appear in the energy gap and cross the Fermi level to accommodate the delocalized excess charge. The uniqueness of the MSS in electrides is that such materials have a topologically different charge density distribution in the bulk from the outside of the surface. The unavoidable delocalization near the surface will generate mobile charge carriers perpendicular to the ACP. It is necessary to point out that the MSS in electrides are universal, and not limited to empty Wyckoff positions and special cleavage planes like in OAI[9].

To better understand the robustness of the electron delocalization and MSS against different surface terminations, an additional Na-A layer is introduced to the top surface of Fig. 2(c). This Na-33-layer model has a mirror symmetry between the top and bottom surface, thus they have the same charge distribution, as confirmed in Figs. 2(c) and 3(b). Now the MSS become doubly degenerated as shown in Fig. 3(a). The wave function in real space is plotted in Fig. 3(c) for the top surface, which gives more details than ELF. It is evident that the electrons do not localize to the atoms, but rather appear as floating states on the surface, being consistent with the Type I scenario as sketched in Fig. 1 at $\lambda = \lambda^*$. We further allow the added Na-A atoms in the Na-33 slab model to relax in $z$ direction. The results are shown in Fig. 3(e). Now the charge state of ISQ4 is decreased from 1.23e to 0e. This again proves that it belongs to the Type I case in Fig. 1 such that on the outside of the surface with $\lambda \to 1$, one must have $\Delta_{l\lambda}^{off} \to 0$ and $n_{l\lambda} \to n_{l\lambda}^{atom}$. The wave function corresponding to the red band is very spreading, as shown in Fig. 3(f).

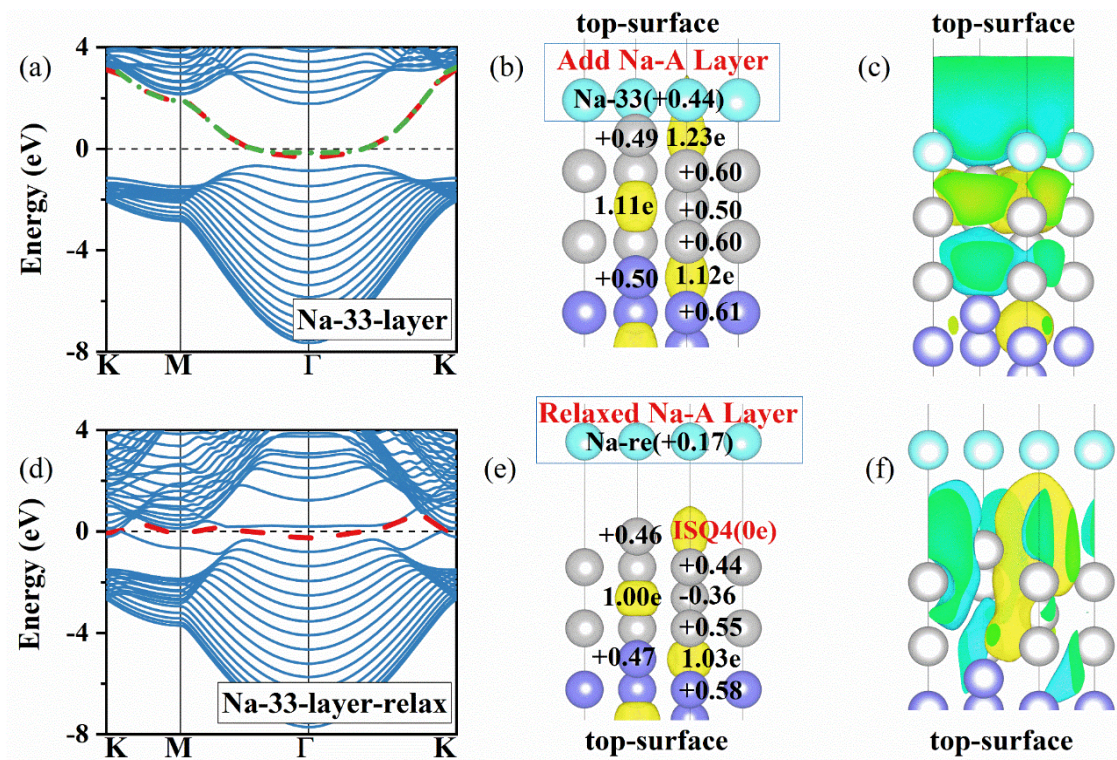

**Figure 3**. (Color online) (a) Band structure of the 8.25-unit-thick slab (33 atomic layers) with an additional Na-A layer; (b) ELF and charge states of the top surface in the Na-33-layer slab model, (isosurface=0.75); (c) The wave function at Γ point shown for the selected red band (isosurface=1×10⁻⁶). Counterpart results for the relaxed Na-33-layer slab model are shown in (d-f), respectively.

The robustness of the MSS against disturbances by different electronegative layer of elements (here only the cases of He, O, and F that cover the typical spectra of electronegativity are displayed for clarity, respectively) are also illustrated in Fig.4. The charge state of ISQ4 increases from 0.63e in the pure Na-hP4 slab to 0.75e when an additional layer of He is added (denoted as He-33-layer). Similar to the case in pure Na-hP4 slab model, the wave function corresponding to the red band indicates that the MSS are mainly contributed by ISQ4, as shown in Fig. 4(c). When the added layer has strong electronegativity, like oxygen and fluorine (denoted as O-33-layer and F-33-layer, respectively), the Bader charge of ISQ4 decreases to 0e. This change is a consequence of the strong interaction between the interface and the localized anionic electrons. In all cases, the MSS is preserved, with ISQ gradually delocalizes and dissipates when approaching the interface.

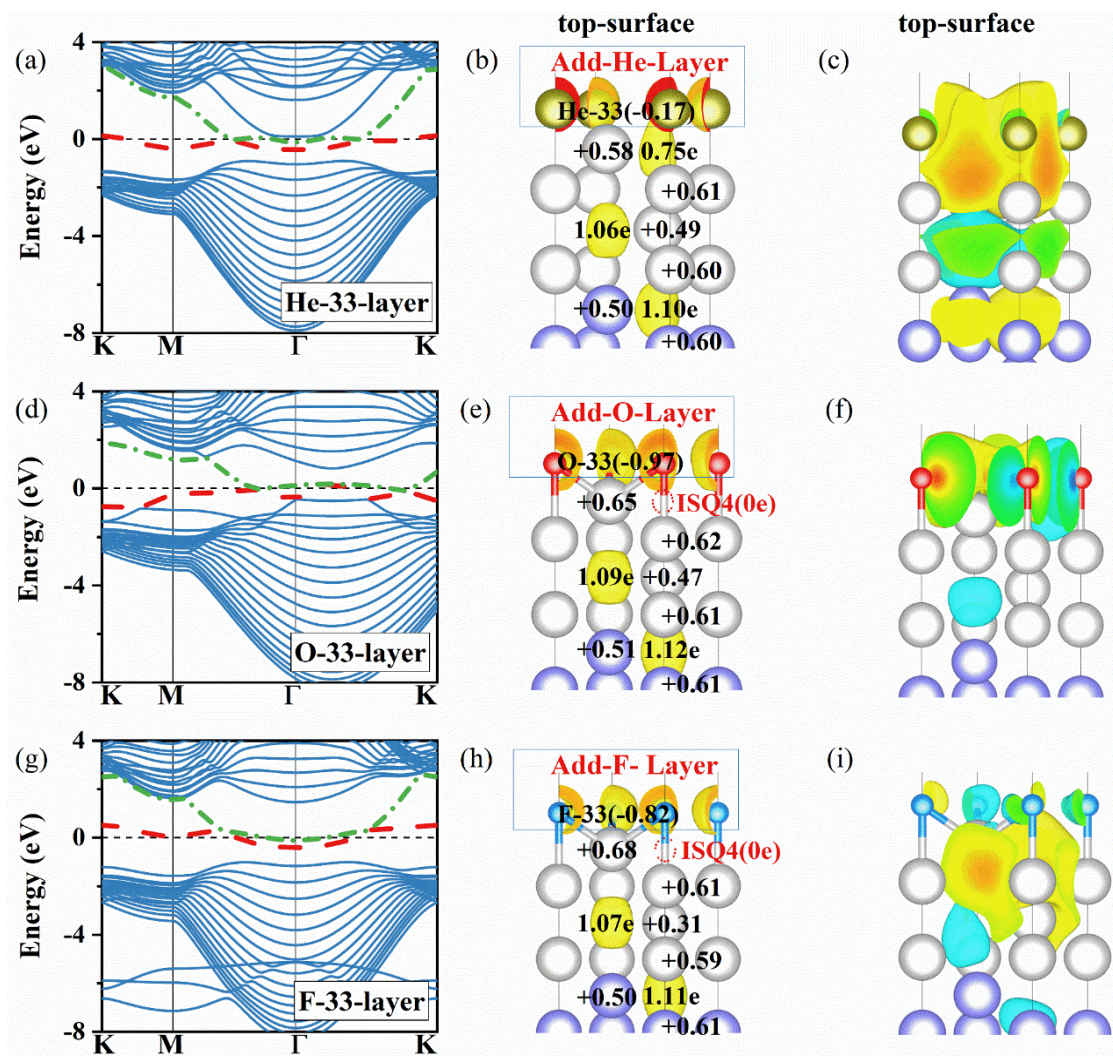

**Figure 4.** (Color online) (a) Band structure of the 8.25-unit-thick slab (33 atomic layers) with an additional He layer; (b) ELF and charge states of the top surface in the He-33-layer slab model, (isosurface=0.80); (c) The wave function at $\Gamma$ point shown for the selected red band (isosurface=1×10⁻⁶). Counterpart results for the O-33-layer (additional oxygen layer added) and F-

33-layer (additional fluorine layer added) slab model are shown in (d)-(f) and (g)-(i), respectively.

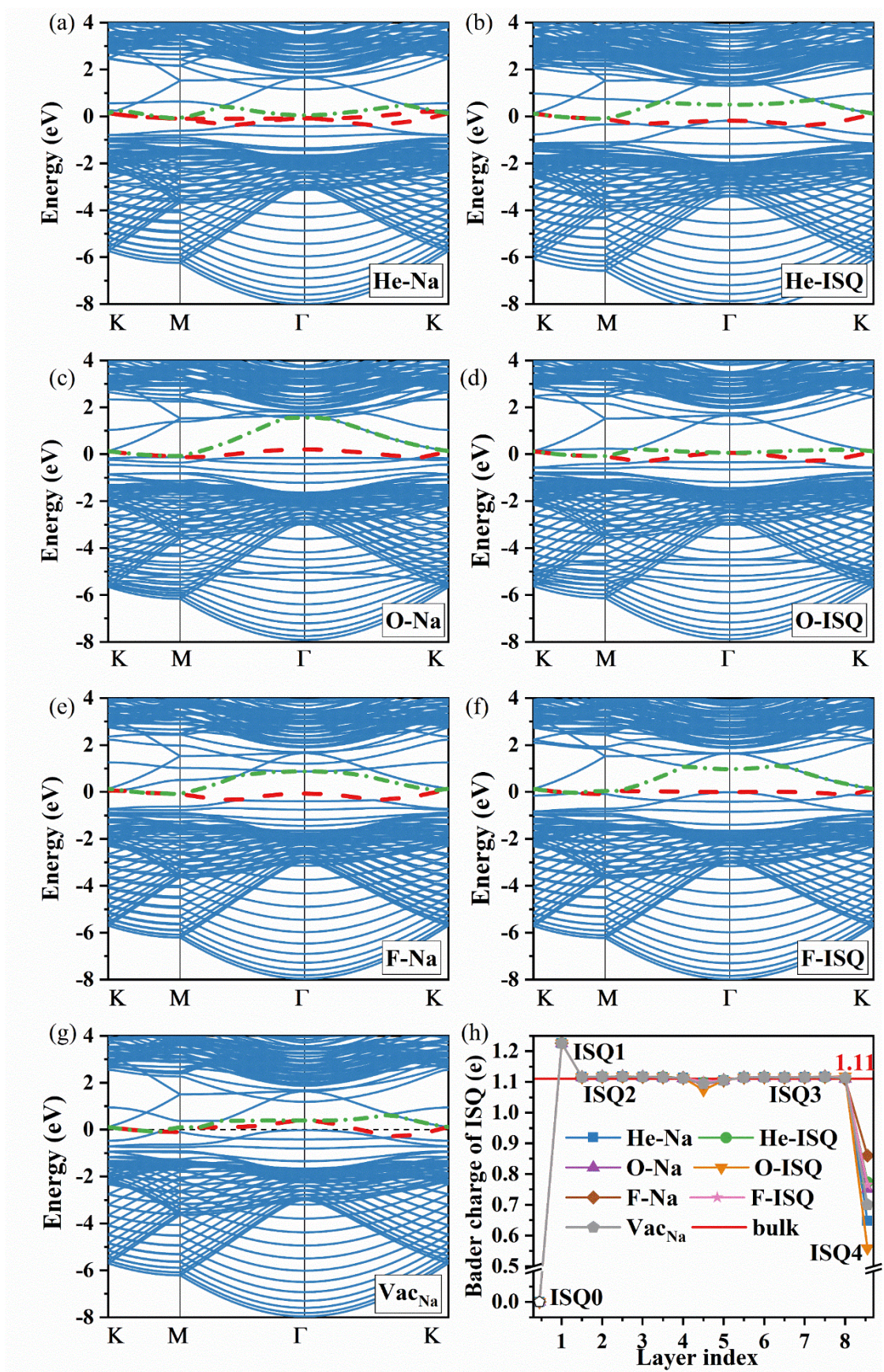

**Figure 5**. (Color online) (a-g) Band structure of the 2×2×1 supercell (128 atoms) of the slab model with impurities or defects included, corresponding to He-Na, He-ISQ, O-Na, O-ISQ, F-Na, F-ISQ, and $Vac_{Na}$, respectively; (h) Variation of the Bader charge of ISQs in the supercell of the slab model, with defects or impurities are introduced to the top surface.

The robustness of the MSS against point defects on the surface is also verified. In Fig. 5(a-g), we displayed the calculated band structure of slab models in which substitutional defects with one Na atom on the surface is replaced by an atom of He, O, or F (labeled as He-Na, O-Na, and F-Na), interstitial defects with He, O, or F atom occupying the ISQ site on the surface (labeled as He-ISQ, O-ISQ, and F-ISQ), and a vacancy defect ($Vac_{Na}$) in a Na-hP4-slab supercell (2×2×1, 128 atoms) are considered, respectively. The corresponding defect concentration of 25% on the surface is high enough to destroy any accident MSS. Even so, it is evident that the surface states are preserved and always locate between the conduction and valence bands, and cross the Fermi level. Figure 5(h) plots the variation of the charge state of ISQs near the defect-doping surface. All cases show a gradual dispersion of ISQ, as well as the charge delocalization. This confirms the prediction of the adiabatic connection.

**3.2 Confirmation of MSS in $Ca_3Pb$, $Cs_3O$, $Y_2C$, and $Sc_2C$**

These four compounds are topologically non-trivial in standard theory according to the Materials Flatband Database[9]. Here we are to demonstrate that their MSS also can be interpreted by the charge delocalization. For example, the 0D electride $Ca_3Pb$ is a semimetal with a topologically nontrivial band structure due to loosely bounded conducting electrons[18]. Following the ACP, we did not find any perceptible displacement of the localized ISQ. Rather, it is dissipated, with the localized charge value of 0.73e in the bulk reduced to 0e at the top surface as shown in Fig. S5(c). The observed delocalization process of $\Delta_{l\lambda}^{off}$ at $\lambda = \lambda^*$ enhances the metallization of the surface, and induces new surface bands at the Fermi level [Fig. S5(d) in the Supporting Information (SI)][62].

Besides 0D electrides, MSS can also be induced in higher dimensional electrides. For example, $Cs_3O$ has been experimentally synthesized and theoretically identified as a 1D topological electride with nontrivial band topology[19]. This material has a large cavity space and thus can be used for various applications such as gas storage, ion transport, and metal intercalation. We have traced the adiabatic variation of the charge density distribution in $Cs_3O$-slab and confirmed that the MSS come from the delocalized ISQ electrons on the surface, as shown in Fig. S6 of the SI.

Special MSS were also reported in 2D electrides, like topological materials $Y_2C$ and $Sc_2C$. By using a 15-layer slab model of $Y_2C$, Huang[20] identified a new surface band originated from the dispersive and floating surface electrons. By using a 15-layer slab of $Sc_2C$, Hirayama[21] verified new surface states in the bulk energy gap originated from the hybridization between the floating states and Sc-3$d$ orbitals. These results confirmed that MSS can always be induced in any dimensional electrides, and all of them obey the Type I mechanism. Above mentioned analysis can be straightforwardly extended to stable semiconducting electrides at ambient pressure, $e.g.$, the $Sc_2C$ with an appreciable band gap. We find that it also supports universal and robust MSS (see

SI).

With these typical but diverse examples, we are able to conclude that a real space topological change in the localized charge density distribution unavoidably leads to a global charge delocalization that accompanied with MSS.

## 3.3 Prediction of MSS with record bulk gap

From above observations, the MSS in electrides are quite general. Following the adiabatic connection principle, we can further predict the universal MSS in other electrides with large bulk gap.

The oP8 phase of ultra-compressed sodium is insulating with a wide band gap at 2 TPa[63]. It is topologically trivial according to standard TBT or TQC[9]. Our calculated results are shown in Fig. 6, the band gap is 6.13 eV, in which a strong $p{\to}d$ electron transition is observed [Fig. 6(a)]. The insulating character arises from strong *s-p-d* multicenter hybridization. According to the classification of OAI[9], Na-oP8 cannot be categorized as OAI. In this sense, one should not expect the existence of MSS. With adiabatic connection, we trace the evolution of local charge density in real space by using an 8-unit layer slab model. Within the bulk, the ISQ with a charge state of 1.24e is localized in the interstices. The valence states of the Na-A and Na-B atoms are +0.58 and +0.66, respectively. When going from the interior of the bulk to the surface, the Bader charge of ISQ1 and ISQ4 are reduced from 1.24e to 1.05e. Namely, the interstitial localized charge is delocalized and becomes dispersive at the surface of the slab [Fig. 6(c)]. We also observe MSS across the Fermi level [Fig. 6(d)], and it is mainly contributed from the ISQ-*s* state. As demonstrated in Fig. 6(d-f), the two-fold degenerated green bands are attributed to the delocalized ISQ1 and ISQ4, whereas ISQ2 and ISQ3 contribute to the red surface bands. It is noteworthy that there is no perceptible displacement of ISQ as a whole when approaching the surface. This again supports the Type I mechanism in Fig. 1. The robustness of these MSS against cleavage planes, lattice terminations, chemical environment of interface, and defects and impurities are also verified, like that done for Na-hP4. Similar predictions are also confirmed in other alkaline earth metals (such as Ca, see Ref.[64]).

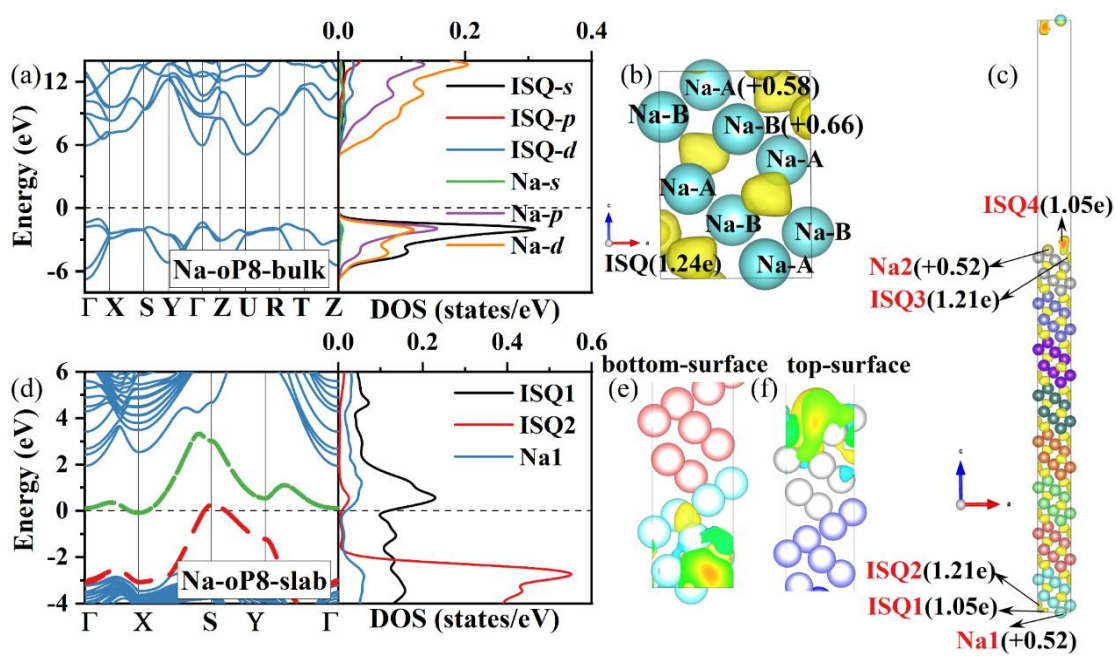

**Figure 6.** (Color online) (a) Band structure (left panel) and PDOS (right panel) of Na-oP8-bulk at 2 TPa; (b) ELF of Na-oP8-bulk, (isosurface=0.75); (c) ELF of Na-oP8-slab, (isosurface=0.75); (d) Band structure (left panel) and PDOS (right panel) of Na-oP8-slab; (e-f) The wave function at the S point corresponding to the red band (dashed line) and the green band (dash-dotted line), respectively (isosurface=$1\times10^{-6}$).

Now we turn to the intermetallic compounds of electrides. At a pressure of 400

GPa, the compound LiNa-oP8 was predicted as an insulating electride with a band gap of 3.67 eV[65]. The Wyckoff sites 4*c* are occupied by both Na and Li atoms. According to Materials Flatband Database[9], LiNa-oP8 is topologically trivial. It is not an OAI, too. As shown in Fig. 7(b), the charge states of bulk LiNa-oP8 are +0.62, +0.52, and -1.14 for Li, Na, and ISQs, respectively. Our calculated band structure of LiNa along the ACP approximated using a slab model with a thickness of 8-unit cell is shown in Fig. 7(d). The characteristic is similar to the high-pressure electride phase Na-oP8. We find that at the surfaces of the slab model, ISQ1 and ISQ4 are partially dissipated, with their charge states being reduced to 0.84e. The excess electrons are pulled into the inside of the surface, increasing the charge state of the adjacent ISQ2 and ISQ3 to 1.25e. This delocalization of ISQs gives rise to mobile carriers and leads to MSS. In Fig. 7(d-f), the red bands contributed by the dispersive ISQ1 and ISQ4 are degenerated. The green bands are mainly from the equivalent ISQ2 and ISQ3.

For all electrides, we find that the delocalization of the interstitial electrons is the main reason for the surface insulator-metal transition. Since the MSS in electrides are independent of the cleavage planes, lattice terminations, and impurities or defects, as well as the chemical environment of the interface or surface, they are universal and robust. The real space distribution of carriers in them has the same topology as TI, thus forms a de facto real space topological phase. Many of them, however, is trivial according to the standard theory. Therefore, it is a new type of topological ordering in real space that is outside the realm of the standard topological theories, such as the TBT in reciprocal space, or the TQC in real space.

The emergence of MSS in this system is described by the global but heterogenous charge delocalization induced by the topology transition of the localized charge distribution from the interstice-centered within the bulk to the atom-centered limit outside of the surface (or interface) along the ACP. Equivalently, the same phenomenon also can be understood in a chemical picture, in which ISQ is described as a collective result of non-local multi-center bonding. The presence of lattice termination or surface boundaries, even if several atomic layers away, will break the congruence of the multi-center bonding (make the bond deficit in electrons), thus the occurring of MSS. This is a local viewpoint for the universal and robust MSS in electrides, comparing to the global viewpoint of topology.

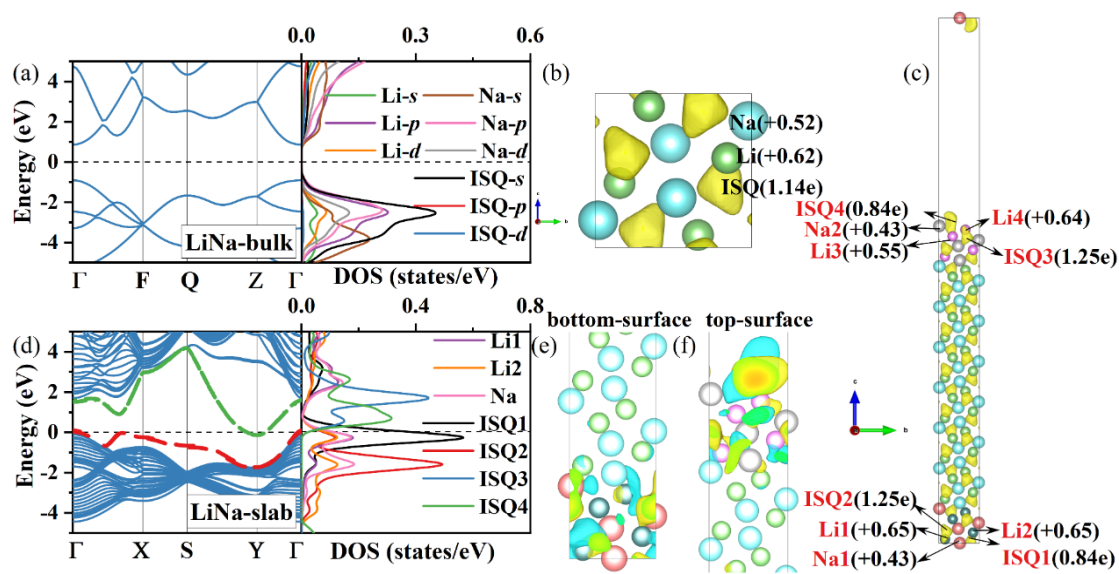

**Figure 7.** (Color online) (a) Band structure (left panel) and PDOS (right panel) of LiNa-bulk at $400$ GPa; (b) ELF of LiNa-bulk, (isosurface=0.75); (c) ELF of LiNa-slab, (isosurface=0.75); (d) Band structure (left panel) and PDOS (right panel) of LiNa-slab; (e-f) The wave function at the S point that corresponds to the red band (dashed line) and the green band (dash-dotted line), respectively (isosurface=1×10⁻⁶).

It is worth mentioning that the charge-core separation in electride could lead to anomalous LA-TA splitting[47]. When electrons are strongly coupled to the lattice vibration, the lattice-driven electronic resonant oscillations could be in-phase or out-of-phase with respect to the acoustic phonons[66]. The latter one has finite energy at zero momentum, and maintains the LA-TA splitting; whereas the former collective excitation (can be called latton) cancels all ionic-induced polarization variations[47], leading to a zero-energy mode at $\Gamma$-point in the reciprocal space, with the frequency $\omega(q) \propto q$ for small momentum $q$, and recovers the usual behavior of the LA branch of the phonon. The implication of this latton excitation with universal MSS in electrides is that an acoustic plasmon resonating with the LA branch of phonon (in addition to the conventional plasmon in metals) could present at the surface or interface of an electride, a phenomenon that might have impact on the electrical or optical properties of the surface/interface[66].

## 4. Conclusions

In summary, electrides are characterized by high localization of charge density to interstitial sites to form ISQs; this could open a band gap with a mechanism different from normal insulators or TIs. By using adiabatic connection formalism, we demonstrated that there is only one energetically possible route for the charge density distribution change when going from the bulk interior of electride adiabatically to the atom-centered limit beyond the surface/interface. After a comprehensive and thorough exploration of all known types of electride (including low-pressure and high-pressure, elementary and multi-component, insulating and semimetallic, as well as 0D, 1D, 2D phases, *etc.*), we discovered that all ISQs dissipate globally, rather than displace as a whole, when approaching the crystalline surface/interface. This real space topology change in the localized charge distribution inevitably leads to a delocalization of the interstitial charge accompanied by the emerging of MSS. We showed that these MSS are stable, being independent of surface terminations, impurities, and defects. Robust and universal MSS in typical insulating electrides, such as Na-oP8 phase, and Li-Na compound, were then presented, even though they all are trivial in standard topology theory. In particular, Na-oP8 phase possesses universal MSS with a huge bulk gap of about 6.13 eV, a size more than 6-fold larger than the highest record of known "wide-gap" TIs.

These findings not only unveiled the limitation of the standard topology theory (in the sense that it fails to detect the hidden real-space topological change in charge distribution), recognized electrides as conceptually novel real-space topological materials, but also illustrated the power of adiabatic connection formalism in the analysis of heterogeneous collective behavior. The resultant conceptual advancement in real space topology opens new opportunity for the design of novel MSS with colossal bulk band gap.

## Conflicts of interest

The authors declare no competing interests.

## Data availability

All data is included in the manuscript and supporting information.

## Code availability

The VASP code used in this paper is commercial software provided by VASP Software GmbH. All others are opensource codes that can be obtained via internet.

## CRediT authorship contribution statement

Dan Wang: Investigation, Methodology, Writing - original draft, Writing - review & editing. Hong X. Song: Methodology, Writing - review & editing, Lei L. Zhang: Methodology, Writing - review & editing. Hao Wang: Methodology, Writing - review & editing. Yi Sun: Methodology. Feng C. Wu: Writing - review & editing. Ying Chen: Software. Xiang R. Chen: Supervision, Project administration. Hua Y. Geng: Idea conceiving, Project design, Writing, Reviewing, and Editing, Supervision, Project administration, Software.

## Supporting Information

MSS of Na-hP4-10$\bar{1}$1 slab; Variation of the maximal values of ELF, the volume of ISQs, the Bader charge of the ISQs, and the charge state of Na-A and Na-B in Na-hP4-0001 slab; ELF of the Na-hP4-0001 slab ($2\times2\times1$ supercell) with impurities and defects; Variation of the maximal values of ELF of Ca$_3$Pb, Cs$_3$O, Na-oP8, and LiNa; MSS of Ca$_3$Pb; MSS of Cs$_3$O; MSS of Sc$_2$C; Detailed data of the structure; Structural stability of high-pressure electrides.

## Acknowledgments

This work was supported by the National Key R&D Program of China under Grant No. 2021YFB3802300, the National Natural Science Foundation of China under Grant No. 12074274, and the NSAF under Grant No. U1730248. Part of the computation was

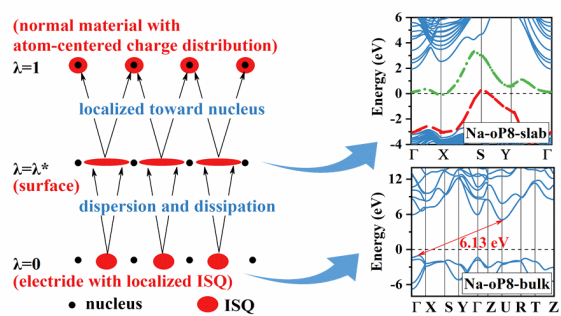

**TOC Graphic**

# Supporting Information for "Universal Metallic Surface States in Electride"

Dan Wang,[1,2] Hongxing Song,[2] Leilei Zhang,[3] Hao Wang,[2] Yi Sun,[2] Fengchao Wu,[2] Ying Chen,[4] Xiangrong Chen,[1*] Hua Y. Geng[2,5*]

[1] *Institute of Atomic and Molecular Physics, College of Physics, Sichuan University, Chengdu 610065, P. R. China;*

[2] *National Key Laboratory of Shock Wave and Detonation Physics, Institute of Fluid Physics, China Academy of Engineering Physics, Mianyang, Sichuan 621900, P. R. China;*

[3] *Institute of Nano-Structured Functional Materials, Huanghe Science and Technology College, Zhengzhou 450063, P. R. China;*

[4] *Fracture and Reliability Research Institute, School of Engineering, Tohoku University, Sendai 980-8579, Japan;*

[5] *HEDPS, Center for Applied Physics and Technology, and College of Engineering, Peking University, Beijing 100871, P. R. China.*

* To whom correspondence should be addressed. E-mail: s102genghy@caep.cn; xrchen@scu.edu.cn.

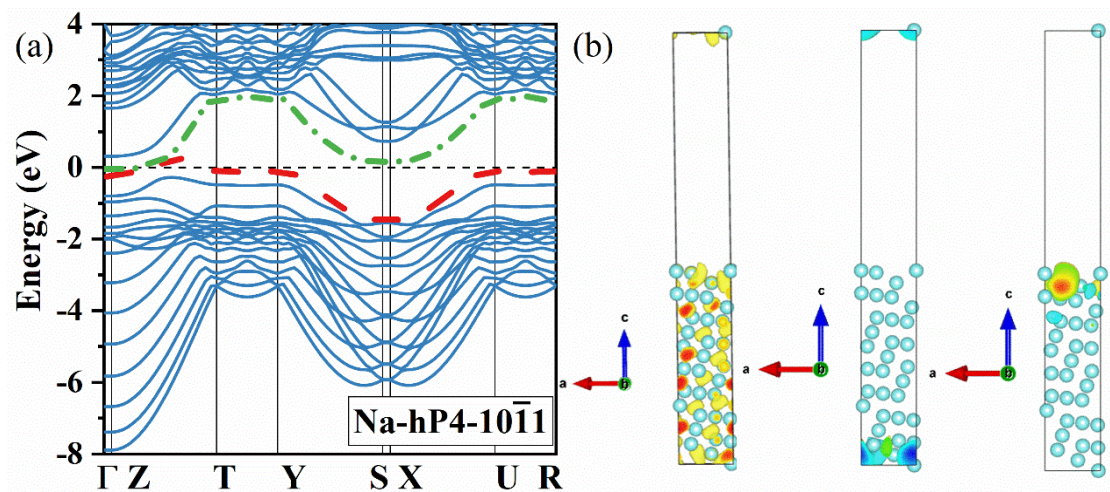

**Figure S1.** (a) Band structure of Na-hP4-slab model whose boundaries are $(10\bar{1}1)$ surface calculated without SOC; (b) Electron localization function (ELF) of Na-hP4-$(10\bar{1}1)$ slab model, (isosurface= 0.75); (c-d) The wave function at the $\Gamma$ point corresponding to the red band (dashed line) and the green band (dash-dotted line) in (a) (isosurface=$1\times10^{-6}$), respectively.

On the Materials Flatband Database, Na-hP4 is an OAI with spin-orbit coupling (SOC), which is indicated by a non-zero $Z$-type real space invariant (RSI) at the Wyckoff position $2d$. As the $2d$ positions are occupied by ISQs and they are not coplanar with any atom on the $(10\bar{1}1)$ plane, it is possible to have a cleavage plane cutting through the Wannier charge center and exhibit the obstructed metallic surface states (MSS). In Fig. S1(a), we confirm the existence of MSS in the Na-hP4-$(10\bar{1}1)$ slab, they are localized in the gap between the valence and conduction bands without SOC. In Fig. S1(b)-(d), ISQs become delocalized and dispersive, making a contribution to the new bands emerging at the Fermi level. In terms of the adiabatic connection concept, there is a significant change in the charge density distribution when approaching the surface, leading to the MSS. Hence the MSS of Na-hP4 will not be affected by the SOC and the crystal itself always exhibits as a topologically trivial insulator.

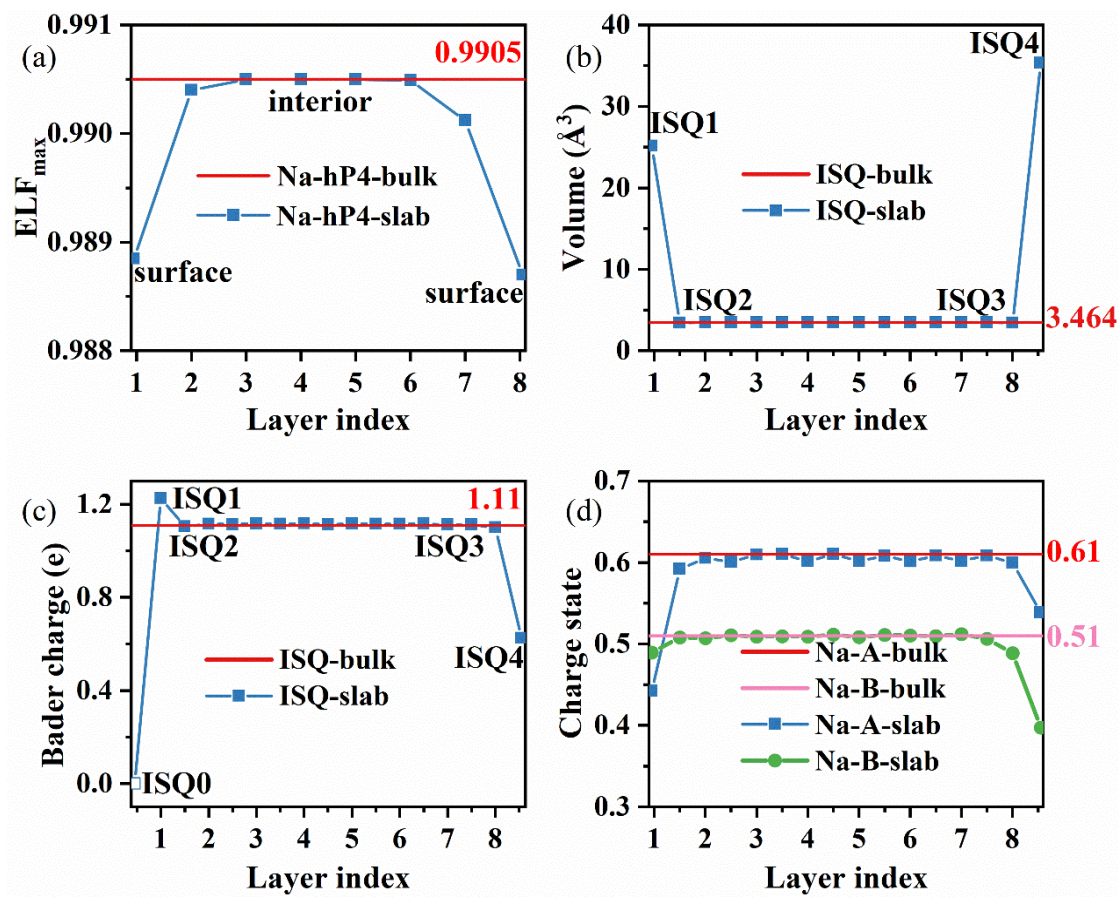

**Figure S2**. (a) Variation of the maximal values of ELF as a function of the layer index obtained in an 8-unit-cell-thick slab (blue line), in comparison with the bulk (red line); (b) Variation of the volume of ISQs in Na-hP4-slab (blue line) and in bulk (red line); (c) Variation of the Bader charge of the ISQs in Na-hP4-slab (blue line) and in bulk (red line); (d) Variation of the charge state of Na-A (blue line) and Na-B (green line) atoms in Na-hP4-slab, comparing with the bulk values [Na-A-bulk (red line) and Na-B-bulk (pink line)], respectively.

The maximal value of electron localization function (ELF) in Na-hP4 is about 0.99, and its variation with the layer index in the slab model is shown in Fig. S2(a). In Fig. S2(b), the volume occupied by the localized interstitial electrons is displayed. It is evident that at surface its value is much large than within the interior, in accordance with the scenario of spreading and dispersion of ISQs. Similarly, the charge state of ISQ1 on the bottom surface is slightly increased [in Fig. S2(c)], driven by the excess electrons from the outside of the surface (where the ISQ0 is completely dissipated). The charge state of ISQ4 on the top surface decreases to 0.63e, leading to charge delocalization. It should be noted that the charge states of ISQs and Na atoms within the interior of the slab are almost identical to the bulk. The calculations with thicker slabs give almost the same results.

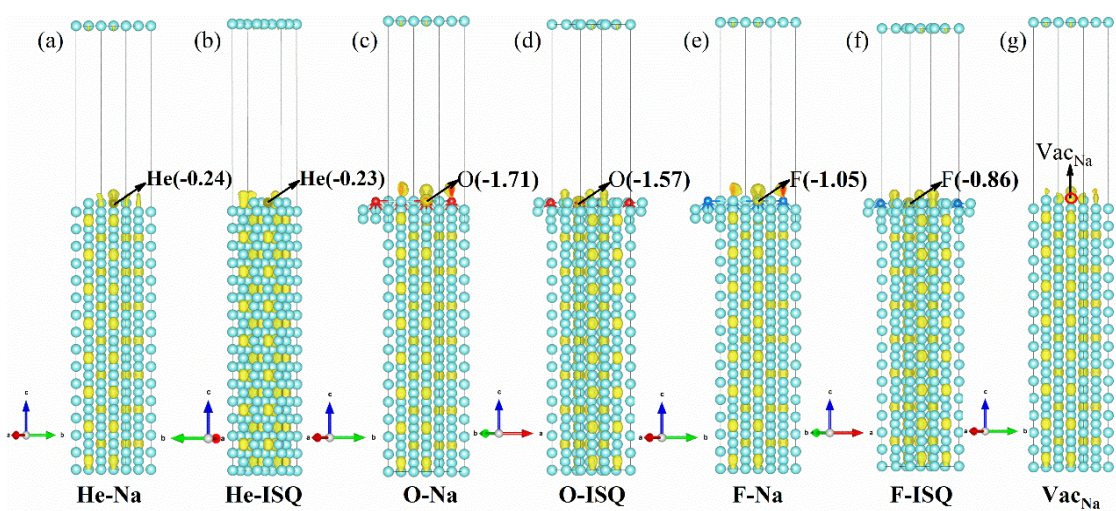

**Figure S3.** ELF (isosurface=0.80) of the 2×2×1 supercell (128 atoms) of the Na-hP4-0001 slab model with 32 atomic layers doping with impurities and various defects, corresponding to the cases of He-Na, He-ISQ, O-Na, O-ISQ, F-Na, F-ISQ and Vac$_{Na}$, respectively.

In order to verify the robustness of the MSS against point defects on the surface, we consider the 2×2×1 supercell (128 atoms) of the Na-hP4-0001 slab model with 32 atomic layers doping with impurities and various defects, labeled as He-Na, He-ISQ, O-Na, O-ISQ, F-Na, F-ISQ and Vac$_{Na}$. The positions of impurities and defects are indicated in Fig. S3, and there is obviously global charge delocalization with the charge transfer from Na atoms to the interstitial atoms and defects.

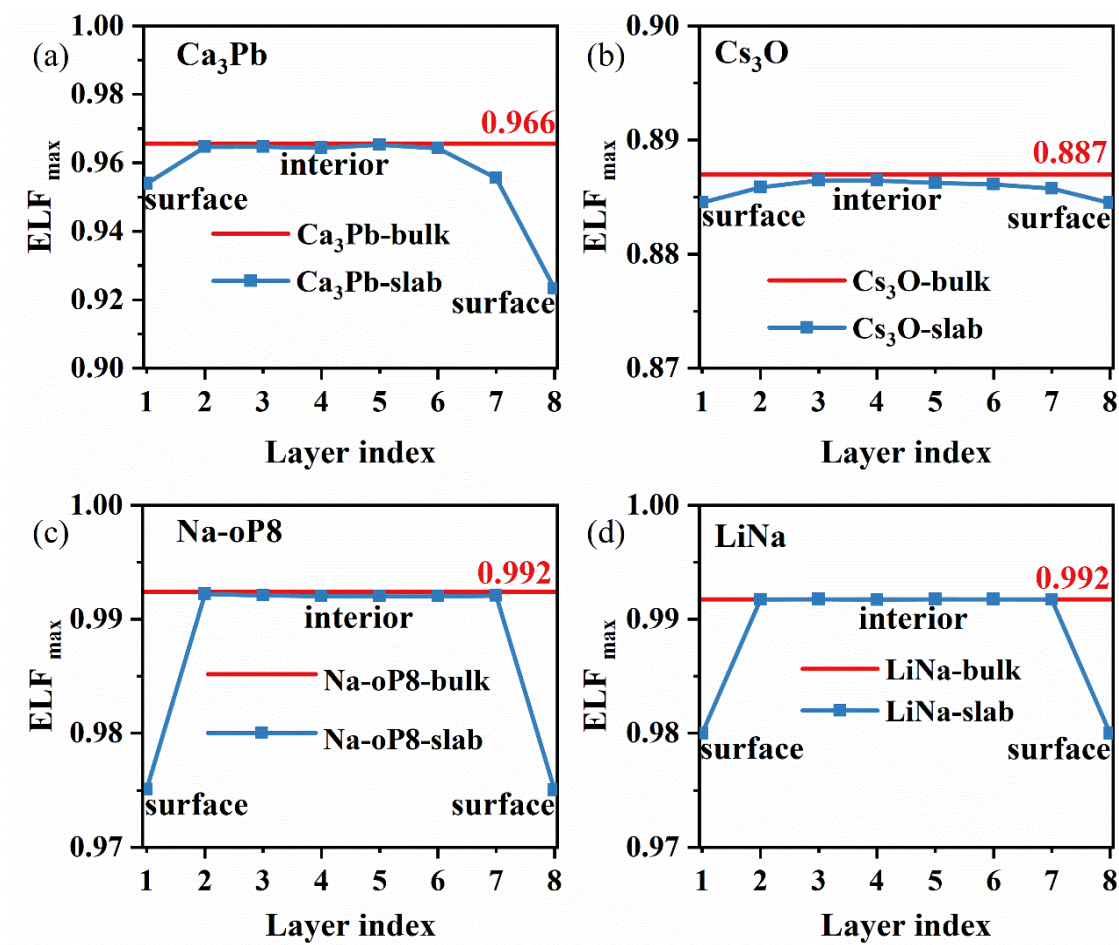

**Figure S4.** Variation of the maximal values of ELF as a function of the layer index obtained in an 8-unit-cell-thick slab (blue line) of $Ca_3Pb$ (a), $Cs_3O$ (b), Na-oP8 (c), and LiNa (d), respectively. The corresponding values of the bulk state are also shown in red lines.

In Fig. S4, the maximal values of ELF in the interior of the slab for $Ca_3Pb$, $Cs_3O$, Na-oP8, and LiNa-oP8 are almost the same as their bulk state, but undergo an obvious reduction when approaching the surface. In the case of the $Ca_3Pb$, the maximal value of ELF on the top surface decreases to 0.92, leading to a charge redistribution at or near the surface. It is evident that the localized ISQs are dissipated gradually, as predicted by the adiabatic connection theory. Similarly, ISQs become delocalized and dispersive when approaching the surface in $Cs_3O$. In our calculation of Na-oP8 and LiNa, the decreased ELF reveals the weakening of the charge localization by the surface, verifying the spreading and dissipation of ISQs. All of these examples refute the Type II mechanism for the evolution of the charge density distribution as illustrated in Fig. 1 of the main text.

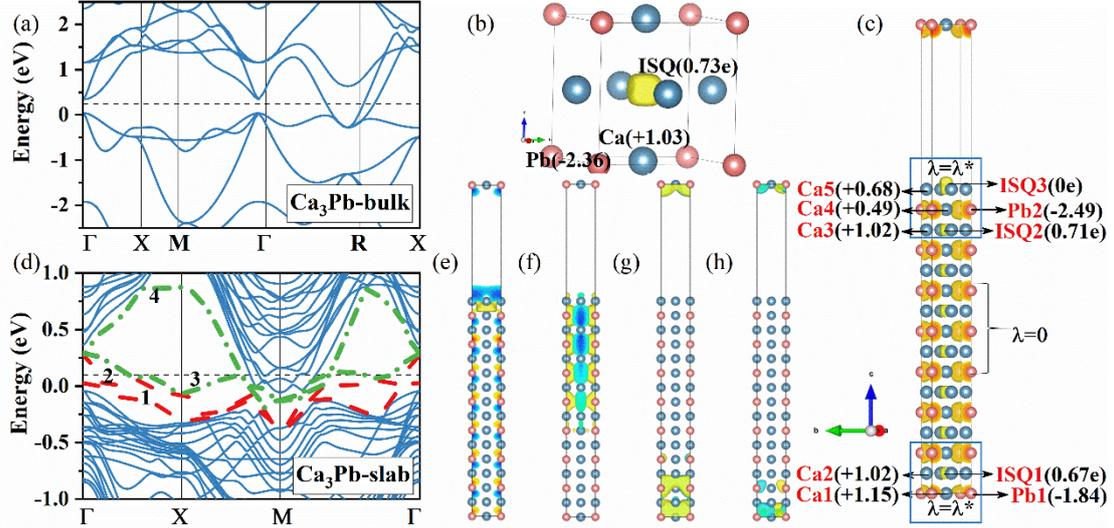

**Figure S5**. (Color online) (a) Band structure of Ca₃Pb-bulk; (b) ELF of Ca₃Pb-bulk, (isosurface=0.75); (c) ELF of Ca₃Pb-slab, (isosurface=0.9); (d) Band structure of Ca₃Pb-slab; (e-f) The wave function at the Γ point corresponding to band 1-4 in (d) (isosurface=2×10⁻⁶).

In Ca₃Pb, a part of electrons are localized to atomic centers, parts are localized to interstitial sites (off-atomic centers), and others are fully delocalized and shared by all atoms. Therefore, the local charge density can be decomposed to $n_{l\lambda} = n_{l\lambda}^{core} + \Delta_{l\lambda}^{off} + \Delta_{l\lambda}^{share}$, in which $\Delta_{l\lambda}^{off}$ corresponds to the charges on localized ISQ. As shown in Fig. S5, when going from the bulk interior to the surface of a slab model, the Bader charges of ISQ1, ISQ2 and ISQ3 all decrease (thus $\Delta_{l\lambda}^{off}$ decreases as well). This leads to the increase of the delocalized charge $\Delta_{l\lambda}^{share}$. Meanwhile some excess electrons go back to the atomic nuclei, and are redistributed between them. Figure S5 (e-h) display the spatial distribution of the wave function of these new surface bands at the Γ point. The eigenstates of the top surface contribute mainly to the red bands (band 1 and 2), while the bottom surface state mainly comes from the green bands (band 3 and 4).

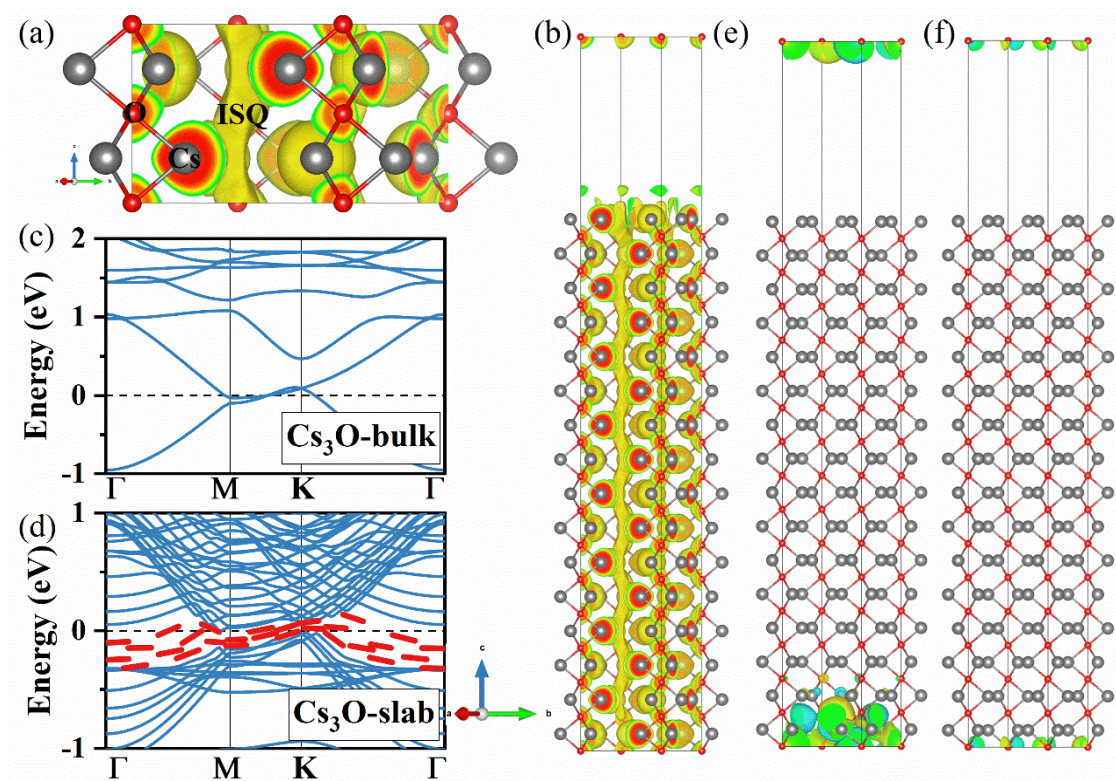

**Figure S6.** (Color online) (a) ELF of $Cs_3O$-bulk (isosurface=0.75); (b) ELF of $Cs_3O$-slab (isosurface=0.75); (c) Band structure of $Cs_3O$-bulk; (d) Band structure of $Cs_3O$-slab; (e-f) The wave function at the $\Gamma$ point corresponding to the red bands in (d) (isosurface=$1 \times 10^{-8}$).

As shown in Fig. S6(a), $Cs_3O$ is a typical one-dimensional electride. The anionic electrons are distributed in the channel voids, forming 1D electron gas. It is evident that the $Cs_3O$-slab with a thickness of 8-unit cell still retains the 1D anionic electrons along the hollow interstitial channel. Nonetheless, the localized 1D ISQ becomes dissipated when approaching the surface, leading to freely moving electrons and inducing new surface bands at or near the Fermi level [see Fig. S6(d-f)].

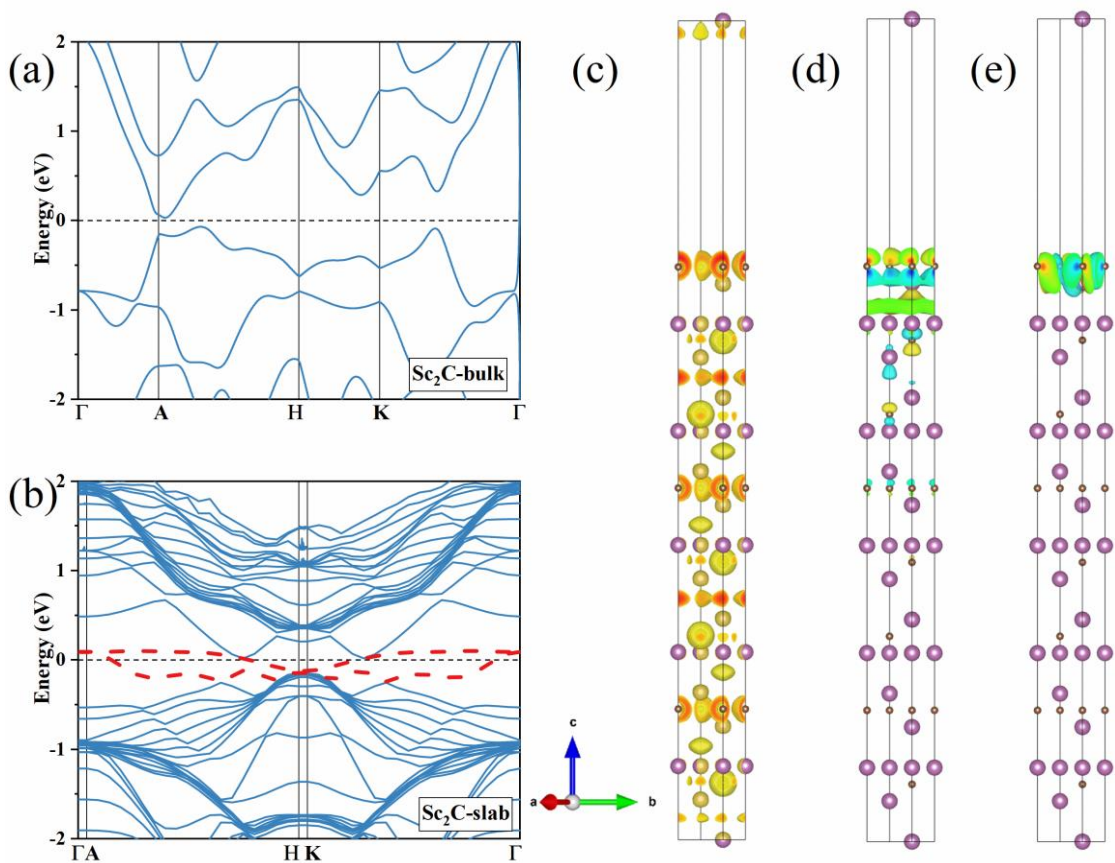

**Figure S7**. (Color online) (a-b) Band structure of Sc$_2$C-bulk and Sc$_2$C-slab, respectively; (c) Electron localization function (ELF) of Sc$_2$C-slab, (isosurface=0.7); (d-e) The wave functions at the $\Gamma$ point corresponding to the red band (dashed line) in (b) (isosurface=1×10$^{-7}$).

The band gap of 2D semiconducting electride Sc$_2$C is 0.31 eV [Fig. S7(a)]. In the band structure of Sc$_2$C-slab, two red bands are formed within the gap of the bulk as shown in Fig. S7(b). The delocalized ISQs at top surface make a contribution to the conducting surface band emerging at the Fermi level, as demonstrated in Fig. S7(d, e). The real space topological change in these localized electrons is the driving force for achieving the MSS of semiconducting electrides at ambient pressure.

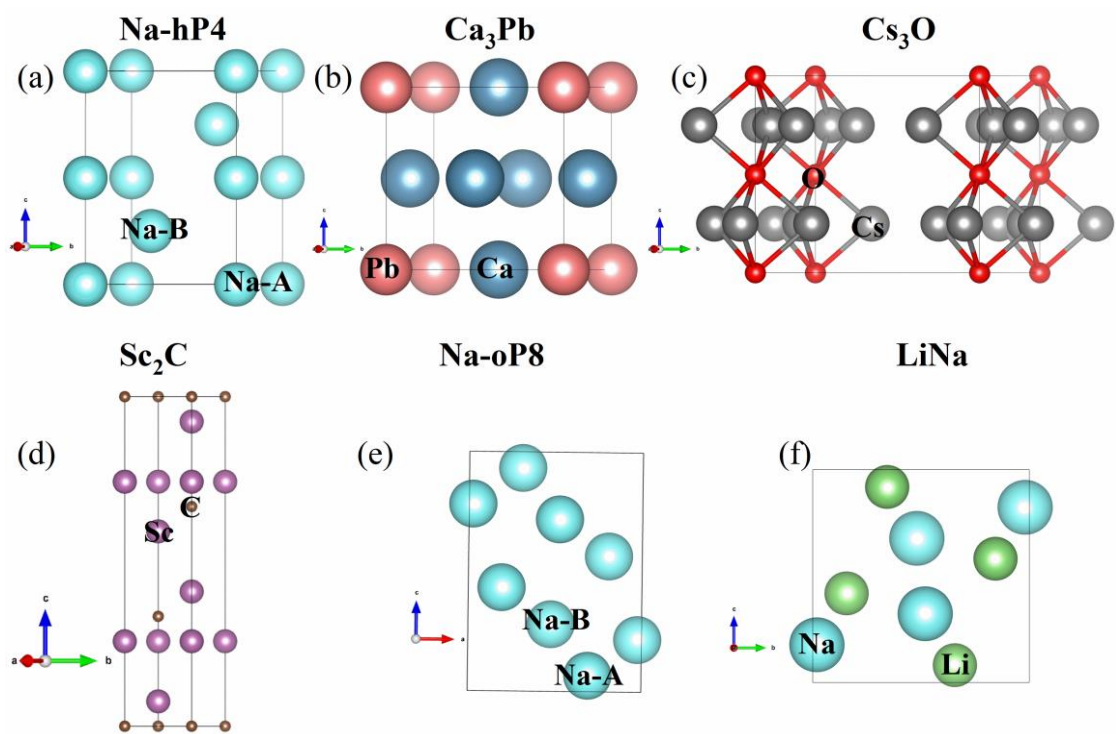

**Figure S8**. (Color online) (a) Structure of Na-hP4-bulk (space group $P6_3/mmc$). Lattice parameters at 320 GPa: $a=b=$2.78 Å and $c=$3.87 Å. There are two inequivalent atomic positions: Na-A atoms locate at Wyckoff site $2a$ (0, 0, 0), Na-B atoms at $2c$ position: (0.333, 0.667, 0.25); (b) Structure of $Ca_3Pb$-bulk (space group $Pm\overline{3}m$). Lattice parameters at ambient condition is $a=b=c=$4.99 Å, the Pb and Ca atoms occupy the $1a$ (0,0,0) and $3c$ (0.5,0.5,0) sites, respectively; (c) Structure of $Cs_3O$-bulk (space group $P6/mmm$). Lattice parameters at ambient condition: $a=b=$8.78 Å and $c=$7.52 Å. Cesium atoms occupy Wyckoff sites $6g$ (0.25, 0, 0.25), oxygen atoms occupy $2b$ position (0, 0, 0), with six Cs atoms connected to a central O atom; (d) Structure of $Sc_2C$ structure (space group $R-3m$). Lattice parameters at ambient pressure: $a=b=$3.62 Å and $c=$17.96 Å. The Wyckoff sites of Sc atoms are $6c$ (0, 0, 0.258), C atoms occupied $3a$ (0, 0, 0); (e)Structure of Na-oP8-bulk (space group $Pnma$). Lattice parameters at 2 TPa: $a=$2.88 Å, $b=$2.12 Å, and $c=$4.0 Å. There are two inequivalent atoms at $4c$ positions: Na-A (0.809, 0.75, 0.567) and Na-B (0.523, 0.25, 0.719); (f) Structure of LiNa-bulk (space group $Pnma$). Lattice parameters at 400 GPa: $a=$2.46 Å, $b=$3.92 Å, and $c=$3.89 Å. The Wyckoff site 4*c* are occupied by Na atoms (0.481, 0.750, 0.177) and Li atoms (−0.156, 0.250, 0.081).

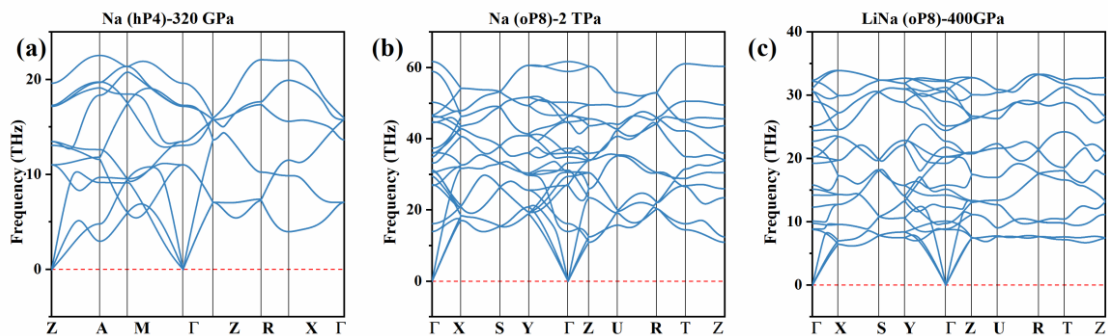

**Figure S9.** (Color online) Calculated phonon spectra of (a) Na-hP4 at 320 GPa, (b) Na-oP8 at 2 TPa, and (c) LiNa at 400 GPa.

In order to estimate the structural stability under high pressures, we further calculated the phonon vibration of these structures. Here, high-pressure electrides are found to be dynamically stable in their respective pressure, as the calculated phonon spectra do not show any imaginary frequency modes.